\documentstyle[amssymb,aps,prl,epsfig]{revtex} 
 
\begin{document} 
\draft     
 
 
\title{Electron momentum distribution of a single mobile hole in the
$t$-$J$ model}  
\author{Anton Ram\v sak} 
\address{J. Stefan Institute,  SI-1000 Ljubljana, Slovenia \\ 
Faculty of Mathematics and Physics, University of Ljubljana, 
SI-1000 Ljubljana, Slovenia 
}
\author{Igor Sega} 
\address{J. Stefan Institute,  SI-1000 Ljubljana, Slovenia 
} 

\date{\today} 
\maketitle 
                
\begin{abstract} 
\widetext 
\smallskip 
We investigate the electron momentum distribution function (EMDF) for
the two-dimensional $t$-$J$ model. The results are based on the
self-consistent Born approximation (SCBA) for the self-energy and the
wave function. In the Ising limit of the model we give the results in
a closed form, in the Heisenberg limit the results are obtained
numerically.  An anomalous momentum dependence of EMDF is found and
the anomaly is in the lowest order in number of magnons expressed
analitycally. We interpret the anomaly as a fingerprint of an emerging
large Fermi surface coexisting with hole pockets.
\end{abstract}
\baselineskip = 2\baselineskip  
\vskip 1 cm

The electron momentum distribution function $n_{\bf k}=\langle
\Psi_{{\bf k}_0}|\sum_\sigma c_{{\bf k},\sigma}^\dagger c_{{\bf
k},\sigma} |\Psi_{{\bf k}_0} \rangle$ is the key quantity for
resolving the structure of the Fermi surface in 
cuprates \cite{marshall96}.  Here we
study the EMDF for $|\Psi_{{\bf k}_0} \rangle$ which represents a
weakly doped antiferromagnet (AFM), i.e., it is the ground state (GS)
wave function of a planar AFM with one hole and with the total
momentum ${\bf k}_0$. In the present work we investigate the
low-energy physics of the CuO$_2$ planes in cuprates within the
framework of the standard $t$-$J$ model
\begin{equation}
H= -t \sum_{<ij>\sigma}  \bigl(
{\tilde c}_{i,\sigma}^\dagger
{\tilde c}_{j,\sigma} + \mbox{H.c.} \bigr)+ 
J \sum_{<ij>}
\bigl[ S_i^z S_j^z+\frac{\gamma}{2}(S_i^+ S_j^- + S_i^- S_j^+ ) \bigr],
\label{tj}
\end{equation}
where
${\tilde c}_{i,\sigma}^\dagger$ (${\tilde c}_{i,\sigma}$) are
electron creation (annihilation) operators acting in a space
forbidding double occupancy on the same site. $S^\alpha_i$ are spin
operators.
Our approach is based on a spinless fermion Schwinger boson
representation of the $t$-$J$ Hamiltonian \cite{sch88} and on the
SCBA for calculating the Green's function  $G_{\bf k}(\omega)$
\cite{sch88,ramsak90,martinez91} and the corresponding wave function
$|\Psi_{{\bf k}} \rangle$ \cite{ramsak93}. 

In general the expectation value $n_{\bf k}$ has to be calculated
numerically.  The Ising limit, $\gamma=0$, is an exception.
The quasi particle is dispersionless with the GS energy 
$\epsilon_{\bf k}=\epsilon_0$, the residue
$Z_{\bf k}=Z_0$ and the
Green's function $G_{\bf k}(\omega)=G_0(\omega)$. Therefore it is
possible to express the required matrix elements in $n_{\bf k}$
analytically and to perform a summation of corresponding non-crossing
contributions to any order $n\to\infty$. The result is
\begin{eqnarray}
n_{\bf k}&=&1-\frac12 Z_0(\delta_{{\bf k} {\bf k}_0}+
\delta_{{\bf k} {\bf k}_0+{\bf Q}})+{1 \over N}
\delta n_{\bf k},\label{dnk}\\
\delta n_{\bf k}&=&{4P \gamma_{\bf k} } 
-4(1-Z_0) \gamma_{\bf k}^2,
\label{ising}
\end{eqnarray}
where $P=\sum_{m=0}^{\infty} \sqrt{A_m A_{m+1}}$ with $A_0=Z_0$,
$A_{m}=A_{m-1}[2 t G_0(\epsilon_0-2 m J)]^2$,
$\sum_{m=0}^\infty A_m=1$ \cite{ramsak93} and $\gamma_{\bf k}=(\cos
k_x +\cos k_y)/2$.  We note that the result
Eqs.~(\ref{dnk},\ref{ising}) exactly fulfills the sum rule 
$\sum_{\bf k} n_{\bf k}=N-1$ and $\delta n_{\bf k}\le1$. In
Eq.~(\ref{dnk}) the only dependence on the GS momentum ${\bf k}_0$
enters through the two delta functions separated with the AFM vector ${\bf
Q}=(\pi,\pi)$. The EMDF $\delta n_{\bf k}$ is determined only
with two parameters, $P$ and $Z_0$,  presented as a function of $J/t$ 
in Fig.~1. Note that $P=1$ and $Z_0=0$ for $J \to 0$, therefore
the result simplifies,
$\delta n_{\bf k}=4 \gamma_{\bf k} (1-\gamma_{\bf k})$.

Now we turn to the Heisenberg model, $\gamma \!\!\to\!\! 1$. Here the
important ingredient is the gap-less magnons with linear dispersion
and a more complex ground state of the planar AFM.  $G_{\bf
k}(\omega)$ is strongly ${\bf k}$-dependent.  The GS is fourfold
degenerate and the results must be averaged over the GS momenta ${\bf
k}_0=(\pm \pi/2,\pm \pi/2)$. To get more insight into the structure of
$\delta n_{\bf k}$, we simplify the wave function by keeping only the
one-magnon contributions.  The leading order contribution to $\delta
n_{\bf k}$ is then
\begin{equation}
\delta n_{\bf k}^{(1)}\!\!=\!-Z_{{\bf k}_0}M_{{\bf k}_0{\bf q}}
G_{{\bf k}_0}(\epsilon_{{\bf k}_0}\!\!-\!\omega_{\bf q})
\bigl[2 u_{\bf
q}+
M_{{\bf k}_0{\bf q}}
G_{{\bf k}_0}(\epsilon_{{\bf k}_0}\!\!-\!\omega_{\bf q})\bigr], \label{dn1}
\end{equation}
\noindent
with ${\bf q}={\bf k}-{\bf k}_0$ [or equivalent in the Brillouin zone
(BZ)], ${\bf v}=t(\sin k_{0x},\sin k_{0y})$, $M_{{\bf k}_0{\bf q}}$
is the hole-magnon coupling and $u_{\bf q}$ is the ususal spin wave
Bogoliubov coefficient \cite{sch88,ramsak90}. The momentum
dependence of the EMDF, contained in Eq.~(\ref{dn1}), essentially
captures well the full numerical solution \cite{nk}.  
A surprising observation
is that the EMDF exhibits in the extreme Heisenberg limit a
discontinuity $\sim Z_{{\bf k}_0} N^{1/2}$ and $\delta n_{\bf k}^{(1)}
\propto -(1+{\rm sign}\, q_x)/q_x$.  We interpret this result as an
indication of an emerging {\it large} Fermi surface at 
discontinuities at {\it points} 
${\bf k}_0$, {\it not lines} in the BZ.

The anomalous structure at ${\bf k}=(\pm \pi/2,\pm \pi/2)$ 
is clearly seen in Fig.~2, where $\delta n_{\bf k}^{(1)}$ is shown
for $Z_{\bf k} t/J \sim 1$ and $\gamma \to 1$. The
Green's function is here approximated with the non-interacting
expression, $G_{{\bf
k}_0}(\omega)\approx-1/\omega$. It should be noted
that $\delta n_{\bf k}^{(1)}$ exhibits at $\gamma=1$ also a (weak) 
singularity ($>1$). However, the $n_{\bf k}$ sum rule is still exactly
satisfied.  In Fig.~2 is for the purpose of presentation $\delta
n_{\bf k}^{(1)}$ truncated to $-6<\delta n_{\bf k}^{(1)}<1$.

In the present work we considered the electron momentum distribution
function for a single hole in AFM and possibly relevant to 
underdoped cuprates. Non-analytic properties
encountered in Eq.~(\ref{dn1}) are an evidence of the emerging {\it
large} Fermi surface at ${\bf k}\sim(\pm \pi/2,\pm \pi/2) $
coexisting, however, with a 'hole pocket' type of a Fermi surface. As
long-range AFM order is destroyed by doping, 'hole-pocket'
contributions should disappear while the singularity in $\delta n_{\bf
k}$ could persist.
We thus 
interpret this result as relevant for the understanding of the electronic
structure found recently with ARPES experiments in uderdoped 
cuprates \cite{marshall96}, where only portions of a large Fermi surface
close to ${\bf k} \sim {\bf k}_0$ were seen. 

%
\noindent
\begin{figure}[htb]   
\center{\epsfig{file=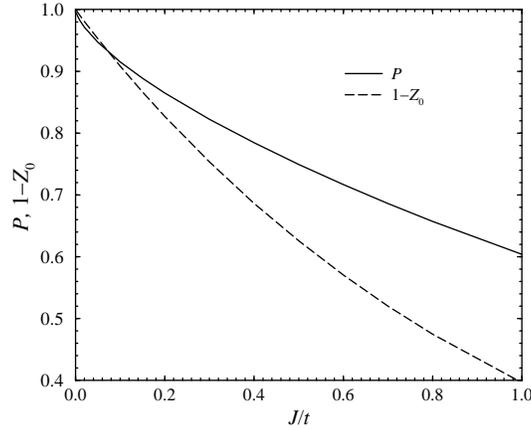,height=80mm,angle=-90}}
\caption{$P=\sum_{m=0}^{\infty} \sqrt{A_m A_{m+1}}$, full line, and
$1-Z_0$, dashed line, determine all momentum dependence of $\delta
n_{\bf k}$ in the SCBA with $\gamma=0$. }
\end{figure}
\noindent
\begin{figure}[htb]     
\center{\epsfig{file=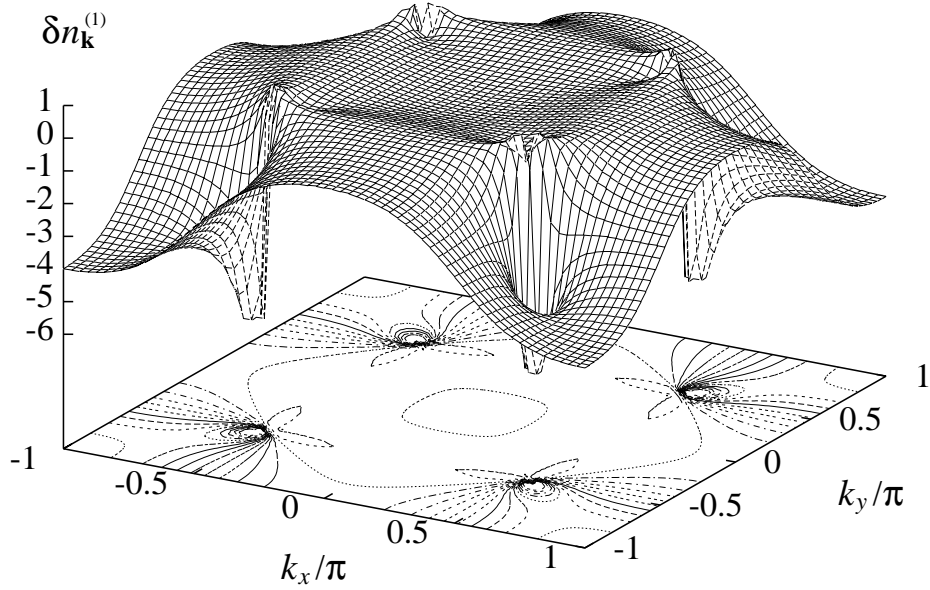,height=120mm,angle=-90}}
\vskip 0.5 cm
\caption{Perturbative result $\delta n_{\bf k}^{(1)}$,
Eq.~(\ref{dn1}), for $Z_{\bf k}
t/J \sim 1$.  }        
\end{figure}

\end{document}